\documentstyle[11pt,paspconf,epsf]{article}

\begin{document}

\title{AGB stars in binaries and their progeny}

\author{A. Jorissen}
\affil{Institut d'Astronomie et d'Astrophysique, Universit\'e Libre de
       Bruxelles C.P. 226, Boulevard du Triomphe, B-1050 Bruxelles, Belgium} 

\begin{abstract}
An AGB star in a binary system is likely to pollute its companion with carbon-
and s-process-rich matter. After the AGB star has faded into an unconspicuous white
dwarf, the polluted companion enters the zoo of stars with chemical
peculiarities. In this paper, the progeny of AGB stars in binary
systems are identified among existing spectroscopic classes (Abell
35-like, binary post-AGB, WIRRing, dwarf Ba and C, subgiant CH, Ba,
CH, S, yellow symbiotics) and their filiation is discussed from the
properties of their eccentricity -- period diagrams.  
\end{abstract}



\section{Introduction}

Although binarity and AGB evolution are in principle disconnected
concepts (a star ought not to be member of a  binary system to evolve 
along the AGB!), a rich world flourishes at their contact. 
Its existence was first suggested by McClure et al. 
(1980; see also McClure 1984 and McClure \& Woodsworth 1990) 
in relation with the discovery of the binary nature of barium
stars. Overabundances of Ba, Sr and other heavy elements produced
by the s-process are observed at the surface of these 
G and K giants. Their origin remained a mystery until McClure suggested that 
the envelope of barium stars may have been polluted by matter accreted from a
former AGB companion (being now a dim white dwarf). 

That scenario raised at first several questions (Is the current companion of
the barium star really a white dwarf? Where are the predicted dwarf barium stars?
What is the mass transfer mode? Is wind accretion efficient
enough? ...) 
that have now largely and satisfactorily been
answered (see the recent reviews by Jorissen \& Van Eck 1998, and Jorissen et al.
1998). The mass-transfer scenario appears to be of great potential to unify
the zoo of red stars with chemical peculiarities, since many such families now fit
within that scenario as reviewed in Sect.~\ref{Sect:unify} 

\section{The mass-transfer scenario and the zoo of peculiar red stars}
\label{Sect:unify}    

\begin{figure} 
\plotone{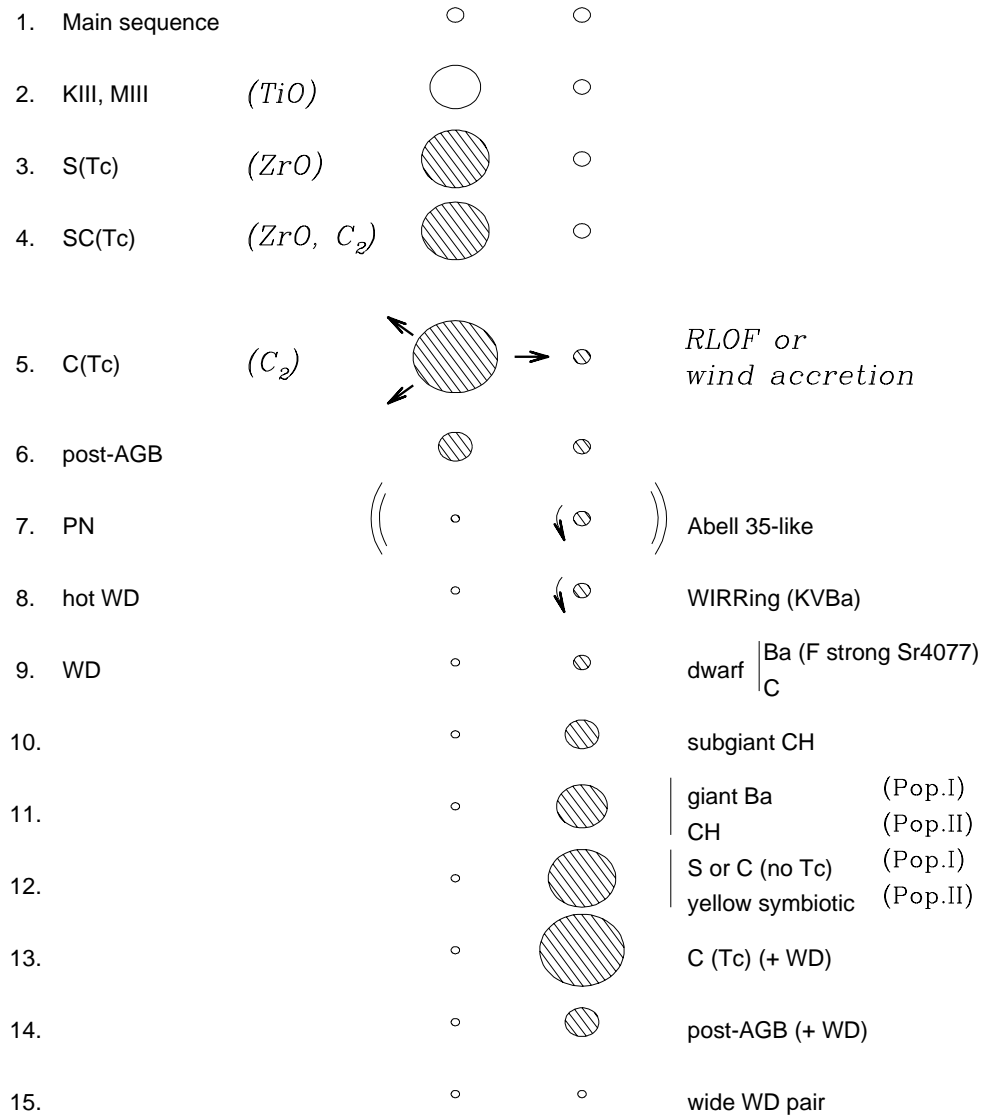}
\caption{\label{Fig:unify}
  Binary evolution involving mass transfer from an AGB star
  (phase 5). Spectral families not requiring binarity (i.e. the normal
  evolutionary sequence for {\it single} stars) are listed in the
  left column, whereas the right column identifies classes of
  PRS {\it requiring binarity}. Stars with excess C and Ba at their
  surface are represented by hatched disks}
\end{figure}

Figure~1 depicts the full evolutionary sequence involving mass-transfer 
from an AGB star in  a binary system (phase 5). As indicated in Sect.~1, it was
originally introduced to account for the chemical peculiarities exhibited by barium
stars (phase 11) which could not be understood in the framework of internal
nucleo\-synthesis in a single star. As reviewed in this section, 
much progress has been made since,
and each phase of this binary evolutionary sequence may now be associated with an
existing class of peculiar red stars (PRS). 
This assignment requires both the typical
chemical signature of pollution from an AGB star (i.e. excess C and Ba) 
to be found in the PRS, and the PRS family to consist exclusively of binary
stars.
\medskip\\
\noindent{\bfseries\slshape Phase 2: Normal giants in binary systems.} 
Catalogues of binary systems involving giant stars 
are provided by Boffin et al. (1993) for
field stars and by Mermilliod (1996) for open-cluster stars. The 
pre-mass-transfer status of these systems is generally not known for sure 
on an individual basis
(since most are SB1 systems, leaving the main-sequence or WD nature of
the companion
uncertain). However, the comparison of their orbital elements with
those of the barium stars
reveals clear differences (further discussed in
Sect.~\ref{Sect:elogP}) which suggest that most of the G-KIII systems are indeed
pre-mass-transfer systems. 
\medskip\\
\noindent{\bfseries\slshape Phases 3-5: S(Tc)-SC(Tc)-C(Tc).} 
Binarity is not required for these PRS exhibiting lines of Tc, an
element with no stable isotopes (Little et al. 1987), as they
represent the normal evolutionary sequence on the AGB (Iben \& Renzini 
1983). 
\medskip\\
\noindent{\bfseries\slshape Phase 5: Mass transfer from a C-rich AGB star.} 
The pollution of the companion occurs at this phase, either
through wind accretion (Boffin \& Jorissen 1988; Han et al. 1995; Theuns et al. 1996;
Karakas et al. 1998) or through Roche lobe overflow (RLOF) (Iben \&
Tutukov 1985; Webbink 1986). 
The efficiency of wind accretion seems to be sufficient to account for
the longest periods observed among PRS (about 30~y, see
Fig.~\ref{Fig:elogP} and Jorissen et al. 1998), and is moreover consistent with the
non-circular nature of the orbits
of those post-mass-transfer systems (Karakas et al. 1998). The occurrence of     
RLOF seems however unavoidable in the systems with circular orbits and periods
shorter than a few hundred days. That question is discussed further in
Sect.~\ref{Sect:elogP}
\medskip\\
\noindent{\bfseries\slshape Phase 6: Binary post-AGB.} 
The binary post-AGB stars identified by Van Winckel et al. (1995 and 
this conference) correspond most likely to this evolutionary phase. 
The mass functions of these SB1 systems do not always allow to
distinguish between a main sequence and a WD companion. Some systems might thus
correspond to phase 14 instead, although the systems with the largest
mass functions
($>0.2$~M$_\odot$) very likely host main sequence companions.
Most (though not all) binary post-AGB stars are very metal-deficient 
(Van Winckel et
al. 1995), and this correlation has been interpreted as the result 
of re-accretion of gas depleted in refractory elements from a stable
circumbinary disk (Waters et al. 1992 and this conference). Because of 
this chemical fractionation, the photospheric chemical content of
these stars does not bear any more the signature of the third
dredge-up. Whether the same (or the complementary) chemical
fractionation alters as well the composition of the matter accreted by
the companion (i.e. the future Ba star) is currently unknown.
\medskip\\
\noindent{\bfseries\slshape Phase 7: Binary nucleus of Planetary Nebulae (Abell
35-like).} 
Among planetary nebulae with binary nuclei, there is a class
consisting of the three objects Abell 35 (BD$-22^\circ$3467 = LW Hya), LoTr 1
and LoTr 5 (HD 112313 = IN Com = 2RE~J1255+255)
whose optical spectra are dominated by late-type (G-K) stars, but whose 
UV spectra indicate the presence of extremely hot 
($> 10^5$~K), hence young, WD companions (see e.g. Bond 
\& Livio 1990). The late-type star is chromospherically active and
rapidly rotating. This rapid rotation is likely to result from some
interaction between the binary components. Since the orbital periods
are still uncertain (Jasniewicz
et al. 1994; Jeffries \& Stevens 1996 and references therein; Gatti et al. 1997), 
the evolutionary history of these
systems cannot unfortunately be identified with certainty: 
close binaries emerging from a common
envelope phase (Bond et al. 1993) or wide binaries whose accreting star has
been spun up by wind accretion (Jeffries \& Stevens 1996)?     
In any case, the detection of moderate overabundances of Ba, Y and Sr 
in the G stars of Abell 35 and LoTr 5 (Th\'evenin \& Jasniewicz 1997;
Gatti et al. 1997 do not confirm, however, that conclusion 
which is based on very few broad lines) 
indicates that these stars have been polluted 
by mass transfer from their former AGB companion. 
The similarity between these systems and the 
WIRRing systems described next may be in favour of a wide binary system.
Furthermore, HST observations of Abell 35 with the Planetary
Camera revealed a measurable extension (0.08 arcsec, translating into
a projected separation of about 18~AU at the minimum estimated distance of 160 
pc) of the nucleus at
300~nm (Gatti et al. 1998), thus ruling out the close-binary hypothesis. 
Another argument favouring wide systems is provided by
HD 128220, a related system consisting of a sdO and a rapidly rotating
G star, which is the only one among those systems having a 
reliably determined period (872~d; Howarth \& Heber 1990).
\medskip\\
\noindent{\bfseries\slshape Phase 8: WIRRing systems (rapidly-rotating KVBa 
+ hot  WD).}    
The class of WIRRing stars (`Wind-induced rapidly rotating')
stars has been introduced by 
Jeffries \& Stevens (1996) to describe a small group of rapidly rotating,
magnetically-active  K dwarfs with hot WD companions uncovered by the ROSAT Wide
Field Camera or EUVE surveys (Kellet et al. 1995; Jeffries et al. 1996). 
From the absence of short-term radial velocity variations, it may be
concluded that these systems have periods of a few months at
least. Moreover, several arguments, based on proper motion, WD cooling
time scale, and lack of photospheric Li, indicate that the rapid rotation of the
K dwarf cannot be ascribed to youth. 
Jeffries \& Stevens (1996) suggest that the K dwarfs in these wide systems were spun
up by the accretion of the wind from their
companion, when the latter was a mass-losing AGB star.    
The possibility of accreting a substantial amount of spin from the companion's
wind has since been confirmed by {\it Smooth Particle
Hydrodynamics} simulations
(Theuns et al. 1996). A clear signature that
mass transfer has been operative
in 2RE~J0357+283 is again provided by the detection of excess barium in
that star (Jeffries \& Smalley 1996).  
 
The class of WIRRing stars may also perhaps include the
barium star HD~165141, since it
shares the properties of RS~CVn and barium systems (Jorissen et al. 1996). The very
long orbital period of the system ($P \sim 4800$~d; Fekel, private
communication) forbids the barium star from
having been spun up by tidal effects as is the case for RS~CVn systems. This
system moreover hosts the hottest WD ($T \sim 35000$~K) among barium systems
(Fekel et al. 1993). The cooling time scale of the WD is about 10~My, shorter
than the magnetic braking time scale of the giant star (about 100~My; see
Theuns et al. 1996). The mass transfer thus occurred recently enough for the
magnetic braking having not yet slowed down the giant star in a substantial way.
HD~165141 is probably one of the very few barium systems where the mass transfer
from the former AGB companion occurred when the accreting star was
already a giant. 
\medskip\\
\noindent{\bfseries\slshape Phase 9: Dwarf Ba/C stars.} 
About a dozen dwarf C stars are now known (Liebert et al. 1979, 1994;
Dearborn et al. 1986; Green et al. 1991, 1992, 1994; Warren et
al. 1993; Heber et al. 1993), and more may be expected from on-going dedicated
surveys (MacConnell 1997). Many dC stars appear to have halo
kinematics. Their binary nature has not been demonstrated in all cases.

Dwarf Ba stars (some with C/O $> 1$) with orbital elements similar to those of the
giant barium stars were identified by North \& Duquennoy (1991, 1992) and North
et al. (1994) among the stars classified as FV strong Sr $\lambda4077$ in the
Michigan spectral survey. The family of CH subgiants identified by
Bond (1974) 
comprises stars near and above the main sequence (Luck \&
Bond 1991). The high occurrence of binaries among subgiant CH stars has
been demonstrated by McClure (1997).   
\medskip\\
\noindent{\bfseries\slshape Phase 12: S(no Tc) or C(no Tc) in Pop.I,
  yellow symbiotics in Pop.II.} 
The identification of the Tc-poor S stars as descendants of the barium stars was
first suggested by Iben \& Renzini (1983) and Little et al. (1987), 
and later confirmed from the similarity
of their orbital elements by Jorissen et al. (1998; see Fig.~\ref{Fig:elogP} and
Sect.~\ref{Sect:elogP}). 
As S(no Tc) stars consist of mass-losing giants and WD companions in
systems with 
orbital
periods similar to those of the symbiotic systems, the question of the relationship
between these two families must be raised. 
This question is actually twofold: (i) Do S(no Tc) 
stars exhibit some symbiotic activity?
(ii) Is the barium syndrome observed among symbiotic systems? 
These questions have been reviewed by Jorissen (1997), and their answers may be
summarized as follows: (i) Some S(no Tc) 
stars indeed exhibit (weak) symbiotic activity (see
below); (ii) no S stars are known among `red' symbiotics (involving M
giants), but all `yellow' symbiotics (involving G or K giants) studied thus far are
halo objects exhibiting the barium syndrome (Schmid 1994; Smith et al. 1996, 1997;
Pereira et al. 1998). As the RGB is shifted towards the blue in low-metallicity
populations, yellow symbiotics may be regarded as the Pop.II analogs
of the S(no Tc) stars
(Smith et al. 1996). The absence of S stars among red symbiotics is more puzzling,
but most likely is a population/metallicity effect: carbon, barium and S stars seem
difficult to form in a high-metallicity environment (Jorissen
\& Boffin 1992; Jorissen \& Van Eck 1998), and
carbon symbiotics are indeed frequent in the Magellanic Clouds
(Miko\l ajewska 1997). A
detailed comparison of the kinematic properties of red symbiotics and
S(no Tc) stars may shed light on that question. 

A recent high-resolution survey of H$\alpha$ emission among the 40 binary
S(no-Tc) stars from  Henize's (1960) sample uncovered only 2 stars with strong, broad
(base
width $\sim$ 400 km~s$^{-1}$) and double-peaked H$\alpha$ emission lines
(Van Eck et al., in preparation) resembling the H$\alpha$
profiles of symbiotic stars (of type S3; Van Winckel et al. 1993) 
and of Abell 35 (Fig.~2 of
Acker \& Jasniewicz 1990).
The distinctive parameter responsible for the symbiotic activity in
these two S stars is currently unknown.

The existence of two kinds of S stars (Tc and no Tc), 
as predicted by the mass-transfer paradigm (phases 3-5 vs. 12 on
Fig.~1), is now well established (Jorissen et al. 1993; Groenewegen
1993; Jorissen \& Van Eck 1998).  Whether the same dichotomy applies
to C-N stars as well is not yet firmly demonstrated, although Barnbaum (1993)
found 16 Tc-poor stars in a sample of 78 C-N stars with Ba excess.
However, their binary nature remains to be established, not an easy
task given the confusion introduced by the envelope pulsation on the
radial-velocity curve (Udry et al. 1998; Hinkle et al., this conference). 
Nevertheless, the recognition of this dichotomy
is important for a correct interpretation of the luminosity
functions of AGB stars in external systems. As shown by Van Eck et al. (1998) from
Hipparcos data, S(no Tc) stars are on average less luminous than S(Tc) stars. They
thus add a low-luminosity tail on the luminosity function of the genuine
AGB S(Tc) stars. If these S(no Tc) [and possibly C(no Tc) stars] are not properly
removed from the luminosity functions observed in external systems, 
they may lead to
erroneous constraints on the luminosity threshold for the occurrence of the third
dredge-up in AGB stars. This danger is best illustrated by the luminosity function
of carbon stars in the SMC obtained by Westerlund et al. (1995), where a formerly
unrecognized low-luminosity tail extends down to $M_{\rm bol} \sim
-2$. This low-luminosity tail precisely matches the luminosity range of S(no-Tc)
stars in the solar
neighbourhood (see Fig.~7 of Van Eck et al. 1998), possibly suggesting that the SMC
may contain an important fraction of binary C(no Tc) stars 
(see also Wood, this conference, for a similar conclusion)!
    
\section{The ($e, \log P$) diagrams of peculiar red stars}
\label{Sect:elogP}

\begin{figure} 
\vspace{-1cm}
\plotone{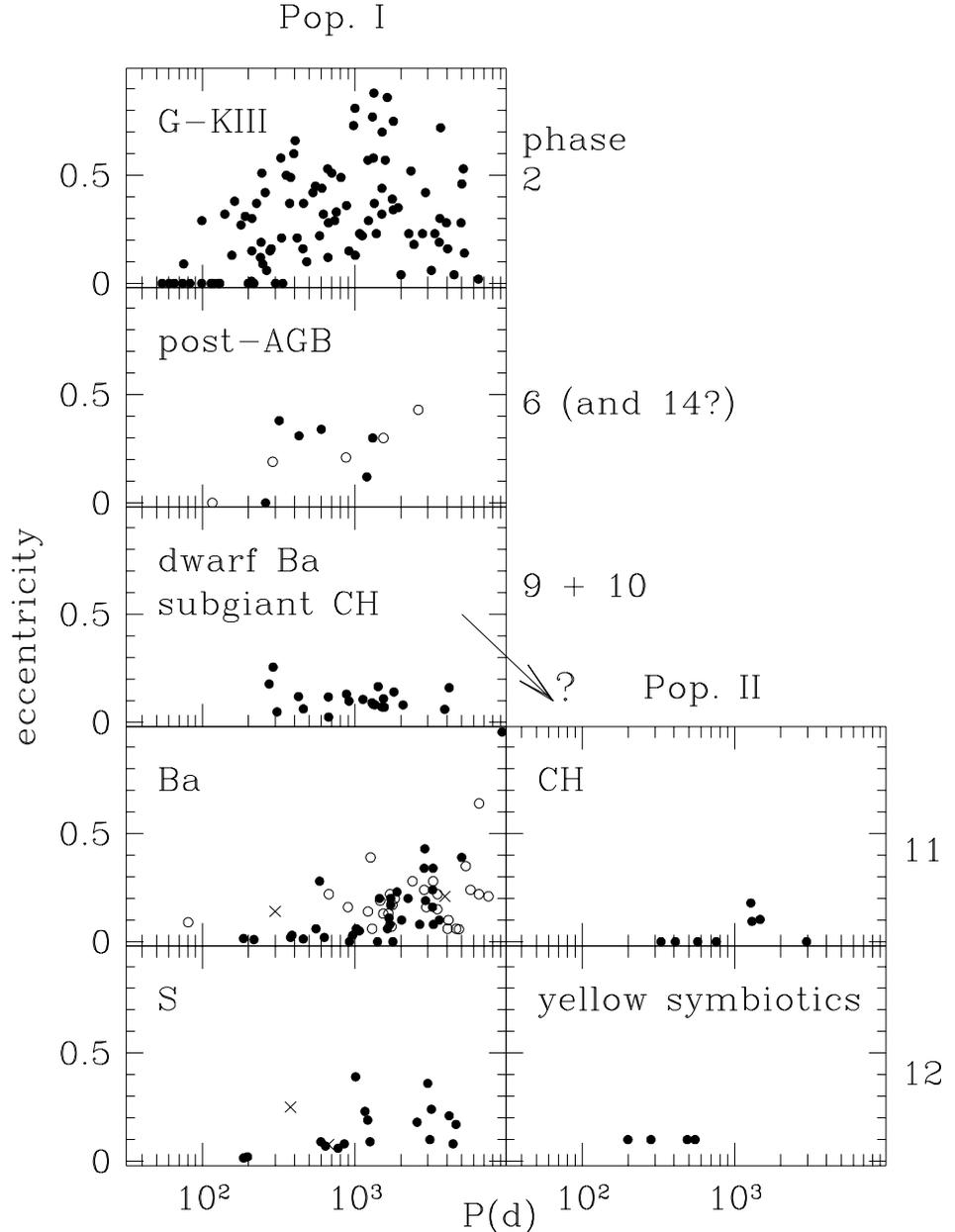}
\vspace{-1cm}
\caption{\label{Fig:elogP}
  Comparison of the $(e, \log P)$ diagrams for the families
  described in Sect.~\ref{Sect:unify}, referred to by the phase label of
  Fig.~\ref{Fig:unify} (written to the right of each panel). Left and right
  columns correspond to Pop.I and Pop.II systems, respectively. 
  Post-AGB systems depleted in refractory
  elements are represented by black dots; systems not exhibiting the
  depletion pattern, or with no abundance data, are represented by
  open dots. The crosses in the
  Ba panel correspond to the two pairs of the triple system
  BD+38$^\circ$118. Ba1,2,3 (i.e. mild Ba) and Ba4,5 (i.e. strong Ba) 
  stars are  represented by open and black dots, respectively.  
  In the S panel, crosses correspond to two systems 
  with unusually large mass functions. As the
  eccentricities of symbiotic systems are generally poorly
  determined, they have all been set to 0.1  
  }
\end{figure}

Figure~\ref{Fig:elogP} compares the ($e, \log P$) diagrams for the
families of binaries described in Sect.~\ref{Sect:unify} 
Orbital elements for {\it G-KIII} stars in open clusters are taken
from Mermilliod (1996), for {\it post-AGB} systems (along with the
sdO+G system HD~128220) from Van Winckel
(this conference) and Howarth \& Heber (1990), 
for {\it dwarf Ba and subgiant CH} stars from North et 
al. (in preparation) and McClure (1997), respectively, for  {\it Ba} and
  {\it S} systems from Jorissen et
al. (1998), and for {\it CH} systems from McClure \& Woodsworth
(1990). 
{\it Yellow symbiotic} systems displayed in Fig.~\ref{Fig:elogP} 
include TX~CVn, BD$-21^\circ3873$, LT~Del (=He2-467) and
  AG~Dra, with orbital elements from various sources 
as listed by Smith et al. (1996). 

The comparison of the $(e, \log P)$ diagrams of these various families 
allows to check their filiation, and to get clues on the mass transfer
process.

The  $(e, \log P)$ diagram of G-K giants in open clusters
differs from all the others
in two clear ways: (i) many cluster giants are found with $P < 200$~d, whereas such
short systems are quite rare among all PRS classes; 
(ii) the average eccentricity at any given period is much smaller for the PRS,
a fact already pointed out by Webbink (1986) for barium stars, but which now turns
out to hold true for all the PRS classes.   These differences between
cluster giants and PRS
are a clear indication that mass transfer (and possibly tidal
interaction) have taken place in the PRS systems. Moreover, the non-zero
eccentricities of most of the PRS systems argue against Roche lobe overflow (RLOF)
as the mass-transfer mode, since this mode requires the mass-loser star to fill its
Roche lobe, a situation where tidal interactions will very rapidly circularize the
orbit. Wind accretion is therefore preferred, as investigated by Boffin \& Jorissen
(1988), Han et al. (1995) and  Karakas et al. (1998). However, wind accretion by
definition requires the system to remain always detached, but as discussed e.g. by
Jorissen et al. (1998) and Karakas et al. (1998), 
this requirement may be difficult to satisfy in systems  with
periods as short as 300~d. RLOF seems thus unavoidable in these
short-period  systems
with circular orbits (at least among Ba systems, but see below). 
That conclusion is troublesome, however, because
RLOF from an AGB star with a deep convective envelope is
generally believed to lead to a common envelope phase (at least when
the mass-losing AGB star is
the more massive component), and to a dramatic orbital shrinkage
(e.g. Paczy\'nski 1976; Meyer and Meyer-Hofmeister 1979; Iben 1995). 
The end product of
such an evolution is a cataclysmic variable with an orbital period
much shorter than
the shortest among PRS systems. The very existence of PRS
systems with periods too short to be accounted for by the wind accretion process
indicates that these systems must have avoided the dramatic fate
outlined above  (see
also Wood, this conference). Several ways out of this channel (which
sometimes does operate, though, since cataclysmic variables {\it do} exist!)
have been sketched by Jorissen \& Van Eck (1998) and Jorissen et
al. (1998; see also Han et al. 1995), but detailed simulations to
confirm  these ideas are badly needed.

The simple picture outlined above of two mass-transfer modes (RLOF for 
the short-period, circular orbits, and wind accretion for the
long-period, eccentric orbits) that seems to emerge from the ($e, \log P$)
diagram of barium systems (after having excluded the triple system
BD$+38^\circ$118), runs into problems when confronted to the $(e, \log 
P$) diagrams of post-AGB, dwarf Ba and subgiant CH systems. The latter 
systems have significantly non-zero eccentricities, even at short
periods! The picture may in fact be complicated by the following
two physical processes: (i) the interaction of a binary system
with a circumbinary disk (a common feature in post-AGB systems; 
see e.g. Waters, this conference) has been shown to pump the eccentricity 
upwards (see e.g. Lubow \& Artymowicz 1992); (ii) tidal
circularization of the closest systems may occur when the Ba and S
giants ascend the RGB, and account for the difference between the
eccentricities of dwarf and giant Ba systems with the {\it shortest} periods.

A mismatch nevertheless remains for the
widest systems, barium stars having larger eccentricities on average
than CH subgiants.
At this point, an observational selection effect against large
eccentricities for CH subgiants with periods close to 
the time span of the monitoring cannot be ruled out as a possible
origin of this mismatch (one long-period, high-eccentricity orbit is
indeed pending in McClure's sample).   
An alternative possibility is that dwarf Ba
and subgiant CH systems evolve into CH stars rather than into
Ba stars.
This second possibility is supported by the similarity of the 
eccentricity and mass function distributions of dwarf Ba stars, CH subgiants and CH
stars (McClure \& Woodsworth 1990; North et al. 1994, 1998;
McClure 1997), and by the often large space
velocities observed for CH subgiants (Bond 1974), which point  
towards subgiant CH stars belonging to a population
intermediate between the halo CH
systems and the old-disk barium systems. 

Finally, a fully satisfactory match is observed between Ba and S systems, which are
distributed identically in the ($e, \log P$) diagram, thus confirming
the hypothesis
that they only differ in terms of effective temperature.    
The same conclusion is reached from the similarity of their mass-function
distributions (Figs.~8 and 10 of Jorissen et al. 1998).

\acknowledgments Pierre North and Hans Van Winckel are thanked for
communicating orbital elements in advance of publication. A.J. is
Research Associate, F.N.R.S. (Belgium).


\begin{references}

\reference Acker A., Jasniewicz J., 1990, \aap~238, 325
\reference Barnbaum C., 1993, \baas~182, \#46.17
\reference Boffin H.M.J., Jorissen A., 1988, \aap~205, 155
\reference Boffin H.M.J., Cerf N., Paulus G., 1993, \aap~271, 125
\reference Bond H.E., 1974, \apj~194, 95
\reference Bond H.E., Livio M., 1990, \apj~355, 568
\reference Bond H.E., Ciardullo R., Meakes M., 1993, in Planetary
Nebulae (IAU Symp. 155), eds. Weinberger R., Acker A., Dordrecht, Kluwer,
p. 397
\reference Dearborn D.S.P., Liebert J., Aaronson M., Dahn C.C., Harrington
R., Mould J., Greenstein J.L., 1986, \apj~300, 314
\reference Fekel F.C., Henry G.W., Busby M.R., Eitter J.J., 1993, \aj~106,
2370
\reference Gatti A.A., Drew J.E., Lumsden S., et al., 1997, \mnras~291, 773
\reference Gatti A.A., Drew J.E., Lumsden S., et al., 1997, in
Planetary Nebulae (IAU Symp. 180), eds. Habing H.J., Lamers
H., Dordrecht, Kluwer, p.105
\reference Gatti A.A., Drew J.E., Oudmaijer R.D., et al., 1998, \mnras, in press
\reference Green P.J., Margon B., MacConnell D.J., 1991, \apj~380, L31
\reference Green P.J., Margon B., Anderson S.F., MacConnell D.J., 1992, \apj~400, 659
\reference Green P.J., Margon B., Anderson S.F., Cook K.H., 1994, \apj~434, 319
\reference Groenewegen M.A.T., 1993, \aap~271, 180
\reference Han Z., Eggleton P.P., Podsiadlowski P., Tout C.A., 1995, \mnras~277, 1443
\reference Heber U., Bade N., Jordan S., Voges W., 1993, \aap~267, L31
\reference Henize K.G., 1960, \aj~65, 491
\reference Howarth I.D., Heber U., 1990, \pasp~102, 912
\reference Iben I.Jr., 1995, Phys. Rep. 250, 1
\reference Iben I.Jr., Renzini A., 1983, \araa~21, 271
\reference Iben I.Jr., Tutukov A.V., 1985, \apjs~58, 661
\reference Jasniewicz G., Acker A., Mauron N., et al., 1994, \aap~286, 211
\reference Jeffries R.D., Smalley B., 1996, \aap~315, L19
\reference Jeffries R.D., Stevens I.R., 1996, \mnras~279, 180
\reference Jeffries R.D., Burleigh M.R., Robb R.M., 1996, \aap~305, L45
\reference Jorissen A., 1997, in Physical Processes in Symbiotic Stars and 
Related Systems, ed. J. Miko\l ajewska, Warsaw, 
Copernicus Foundation for Polish Astronomy, p.135
\reference Jorissen A., Boffin H.M.J., 1992, in  Binaries as
tracers of stellar formation, eds. Duquennoy A., Mayor M. Cambridge
Univ. Press, p. 110
\reference Jorissen A., Frayer D.T., Johnson H.R., et al., 1993,
\aap~271, 463
\reference Jorissen A., Schmitt J.H.M.M., Carquillat J.M., et al.,
1996, 
\aap~306, 467
\reference Jorissen A., Van Eck S., 1998, in The Carbon Star Phenomenon (IAU Symp.
177), ed. R.F. Wing, San Francisco: ASP, in press
\reference Jorissen A., Van Eck S., Mayor M., Udry S., 1998, \aap~332, 877
\reference Karakas A.I., Tout C.A., Lattanzio J., 1998, \mnras, in press 
\reference Kellet B.J., Bromage G.E., Brown A., et al., 1995,
\apj~438, 364
\reference Liebert J., Dahn C.C., Gresham M., Strittmatter P.A., 1979, \apj~233, 226
\reference Liebert J., Schmidt G.D., Lesser M., et al., 1994,\apj~421, 733
\reference Little S.J., Little-Marenin I.R., Hagen Bauer W., 1987, \aj~94, 981
\reference Lubow S.H., Artymowicz P., 1992, in
Binaries as tracers of stellar formation,  A. Duquennoy, M. Mayor (eds.). Cambridge
UP, p.145   
\reference Luck R.E., Bond H.E., 1991, \apjs~77, 515
\reference MacConnell D.J., 1997, Baltic Astronomy 6, 105
\reference McClure R.D., 1984, \pasp~96, 117
\reference McClure R.D., 1997, \pasp~109, 536
\reference McClure R.D., Fletcher J.M., Nemec J.M., 1980, \apjl~238, L35
\reference McClure R.D., Woodsworth, 1990, \apj~352, 709
\reference Mermilliod J.-C., 1996, in The Origins, Evolution, and Destinies of Binary
Stars in Clusters, Milone G.F., Mermilliod J.C., ASP
Conf. Ser. 109, p. 373
\reference Meyer F., Meyer-Hofmeister E., 1979, \aap~78, 167
\reference Miko\l ajewska J., 1997, in Physical Processes in Symbiotic Stars and 
Related Systems, ed. J. Miko\l ajewska, Warsaw, 
Copernicus Foundation for Polish Astronomy, p.3
\reference North P., Duquennoy A., 1991, \aap~244, 335
\reference North P., Duquennoy A., 1992, in Binaries as
tracers of stellar formation, eds. Duquennoy A., Mayor M.
Cambridge Univ. Press, p.202
\reference North P., Berthet S., Lanz T., 1994, \aap~281, 775
\reference North P., Jorissen A., Mayor M., 1998, in 
The Carbon Star Phenomenon (IAU Symp. 177), ed. R.F. Wing, San Francisco: ASP, in press
\reference Paczy\'nski B., 1976, in Structure and Evolution of Close
Binary Systems (IAU Symp. 73), Eggleton P.P., Mitton S., Whelan J.,
Reidel, p. 75
\reference Pereira C.B., Smith V.V., Cunha K., 1998, \aj, in press
\reference Schmid H.-M., 1994, \aap~284, 156
\reference Smith V.V., Cunha K., Jorissen A., Boffin H.M.J., 1996, \aap~315, 179
\reference Smith V.V., Cunha K., Jorissen A., Boffin H.M.J., 1997, \aap~324, 97
\reference Theuns T., Boffin H.M.J., Jorissen A., 1996, \mnras~280,
1264
\reference Th\'evenin F., Jasniewicz G., 1997, \aap~320, 913
\reference Udry S., Jorissen A., Mayor M., Van Eck S., 1998,
\aaps~131, 25
\reference Van Eck S., Jorissen A., Udry S., Mayor M., Pernier B., 1998,
\aap~329, 971
\reference Van Winckel H., Duerbeck H.W., Schwarz H.E., 1993,
\aaps~102, 401
\reference Van Winckel H., Waelkens C., Waters L.B.F.M., 1995,
\aap~293, L25
\reference Warren S.J., Irwin M.J., Evans D.W., et al., 1993, \mnras~261, 185
\reference Waters L.B.F.M., Trams N.R., Waelkens C., 1992, \aap~262, L37
\reference Webbink R.F., 1986, in Critical Observations vs. Physical Models for
Close Binary Systems, eds. K.-C. Leung, D. Zhai, Gordon and Breach, p. 403  
\reference Westerlund B.E., Azzopardi M., Breysacher J., et al. , 1995,
\aap~303, 107

\end{references}
\end{document}